\documentclass{article}

\usepackage{arxiv}

\usepackage[utf8]{inputenc} 
\usepackage[T1]{fontenc}    
\usepackage{hyperref}       
\usepackage{url}            
\usepackage{booktabs}       
\usepackage{amsfonts}       
\usepackage{nicefrac}       
\usepackage{microtype}      
\usepackage{lipsum}		
\usepackage{graphicx}
\usepackage{doi}

\usepackage[tight,footnotesize]{subfigure}
\usepackage{amsfonts}
\usepackage{amsmath}
\usepackage{enumerate}

\usepackage[]{biblatex}
\usepackage{csquotes} 
\title{PCB Transducer Coil Design for a Low-Noise Magnetic Measurement System in Space Missions
}

\author{ Jose A. Vílchez Membrilla   \\
	Departamento Ingeniería de Sistemas y Electrónica,\\
	Universidad de Cádiz\\
	Cádiz, 11519, Spain \\
	\texttt{joseantonio.vilchez@uca.es} \\
	\And
    Ignacio Mateos \\
	Departamento Ingeniería de Sistemas y Electrónica,\\
	Universidad de Cádiz\\
	Cádiz, 11519, Spain \\
	\texttt{ignacio.mateos@uca.es} \\
    \And
    Ángel Quirós-Olozábal \\
	Departamento Ingeniería de Sistemas y Electrónica,\\
	Universidad de Cádiz\\
	Cádiz, 11519, Spain \\
	\texttt{angel.quiros@uca.es} \\
    \And
    Mario F. Pantoja \\
	Departamento de Electromagnetismo,\\
	Universidad de Granada\\
	Granada, 18071, Spain \\
	\texttt{ignacio.mateos@uca.es} \\
    \And
    Clemente Cobos Sánchez \\
	Departamento Ingeniería de Sistemas y Electrónica,\\
	Universidad de Cádiz\\
	Cádiz, 11519, Spain \\
	\texttt{clemente.cobos@uca.es} \\
}

\date{}

\renewcommand{\headeright}{\textcopyright{} [2022] IEEE. Personal use permitted. }
\renewcommand{\undertitle}{\textcopyright{} [2022] IEEE. Personal use permitted. Published version available at: \url{https://doi.org/10.1109/tim.2022.3217843
}}
\renewcommand{\shorttitle}{PCB Transducer Coil}

\hypersetup{
pdftitle={PCB Transducer Coil Design for a Low-Noise Magnetic Measurement System in Space Missions},
pdfsubject={},
pdfauthor={Jose A.~Vilchez},
pdfkeywords={},
}

\begin{document}
\maketitle

\begin{abstract}
	Here a minimum power dissipation coil transducer has been designed to allow in-flight noise characterization of a low-frequency magnetic measurement system. The coil was produced  using PCB technology to provide mechanical stability and easy integration close to the magnetic sensors. The transducer design strategy relies on an inverse boundary element method, which has been efficiently adapted to  produce  optimal  multilayer  PCB coils subjected to different geometrical and performance constraints. The resulting coil transducer was simulated and experimentally validated under realistic conditions where the payload is magnetically shielded from low-frequency fluctuations by using a three-layer cylindrical mu-metal enclosure. The results show that the specific transducer coil proposed efficiently produces stable magnetic conditions for adequate validation of the magnetic diagnostic system while providing reduced power dissipation and optimal mechanical stability.
\end{abstract}


\section{Introduction}

Detection and precise measurements of magnetic fields have become essential tasks in many fields, such as industrial processes, medical applications and space research. A good example of the latter  is  the observation of Gravitational Waves (GWs) \cite{LISAWhitePaper}.

The extreme weakness of GWs, along with the harsh environmental conditions found in space, give rise to a crucial technical problem: it must be possible to distinguish the GWs from undesired non-gravitational noise \cite{Mateos2015}, such as that generated by interplanetary magnetic fields or magnetic sources on the spacecraft itself. This raises the need for precise electronic measurement systems capable of mapping the magnetic forces acting on the spacecraft \cite{MateosQuaGrav}.

 Of all the different sensing techniques employed in magnetic diagnostic systems for space-based GW detectors, anisotropic magnetoresistive (AMR) sensors are preferred to other technologies due to their small size, low magnetic impact on the surroundings, low power consumption and high sensitivity \cite{AMR_uses}. This is the case of initiatives such as MELISA-III (Magnetic Experiment for the Laser Interferometer Space Antenna). \footnote{MELISA-III is part of an In-Orbit demonstration/Validation (IOD/IOV) activity developed at the University of Cádiz (UCA) and selected by the European Commission and the European Space Agency (ESA) under the H2020 Programme \cite{MateosCubeSat}. } MELISA-III aims to validate a compact magnetic measurement system for use in future space-based gravitational wave detectors. The main objective of MELISA-III is the in-flight sub-millihertz noise characterization of a set of magnetoresistive sensors with dedicated electronic noise reduction techniques. With the aim of enhacing the sensitivity and linearity of the magnetic measurements, the core of an AMR magnetic sensor is made up of an all-element varying Wheatstone bridge, i.e., the four elements of the bridge are magnetoresistances.

The excess noise at sub-millihertz frequencies is the critical part of the payload which arises from the intrinsic flicker noise  and the slow environmental temperature drifts coupled with the thermal coefficient of the electronics. Although both noise contributions are dominated by the sensor itself, the latter contribution increases linearly with the magnetic field. Accordingly, the AMR sensor also integrates a compensation coil that is used as part of a closed-loop controller to balance the bridge during operations. This allows improvement of the thermal dependence of the sensor sensitivity by using an integrator that supplies the compensation coil, generating an inverse field to cancel out the field component along the sensor \cite{MateosCubeSat}.

For the in-orbit validation of the MELISA-III experiment, a bias field must be applied by another external transducer coil placed within a mu-metal enclosure to ensure stable magnetic conditions in the harsh Low Earth Orbit (LEO) environment. This external coil will allow low-frequency noise performance  over the whole magnetic measurement range of the sensor, where the figure of merit will be the behavior of the flicker noise, the $1/f$ corner frequency and the noise floor.

 Other examples exist of space missions where on-board coils are needed to inject controlled magnetic fields during in-flight scientific operations \cite{CoilPCB_LPF}-\cite{schuldt_STE-QUEST}, yet none of them have the strict design constraints of the CubeSat missions in terms off mass, volume and power. More precisely, in the case of the MELISA-III experiment the following technical requirements must be taken into account in the design of an ideal transducer coil for the payload:
 
 \begin{enumerate}
  \item[i)] High mechanical stability, especially to ensure that the system will safely survive the launch mechanical environment. 
 \item[ii)] Reduced physical dimensions and easy integration into the electronic magnetic diagnostic system.
 \item[iii)] Proximity to the magnetic sensor, so as to achieve the desired magnetic field with the lowest possible current.
  \item[iv)] Maximum desired magnetic field at the specific location of the sensor core within the integrated circuit.   
 \item[v)]   Minimum resistance in order to reduce Joule heating losses, a fact which may become especially  significant at a small coil wire scale. 
 \end{enumerate} 
  An interesting strategy to meet these requirements is the design of planar coils making use of the different metal layers of the multilayer PCB structure where the magnetic diagnostic circuit is implemented. The use of multilayer PCB technology has several potential advantages, such as the ease of fabrication of smaller transducers and the compact and precise integration of the coil in the circuit assembly \cite{Rogowski2021,Fluxgate_Baschiroto,Noroozi2021,Nireeshma2022,Yan2009}, which can both ameliorate  mechanical stress due to high accelerations/vibrations (compared to other external coil shapes) and reduce the distance between the transducer and the magnetic sensors.

 There is a considerable literature on the problem of coil design in electronics \cite{Zhao2022,Yang2021,Zhao2021_PSO,Ding2022}, nonetheless these approaches are based on heuristic algorithms, restricted to simple coil geometries such as cylindrical or planar shapes and unable to satisfy the technical requirements described in i)-v). 
 On the other hand, coil design problem (also known as magnetic field synthesis) has been widely studied in other fields of engineering such as bio-engineering \cite{Poole_2007,KoponenMinimum} and magnetic accelerators \cite{Koyanagi2015}.
Moreover, among all the approaches developed in recent years to design coils there is an especially successful series of techniques which incorporates the stream function of a current density within an inverse boundary element method (IBEM) \cite{Pissanetzky_1992}. This approach has been efficiently applied to produce magnetic resonance imaging (MRI) gradient coils and transcranial magnetic stimulation (TMS) coils with arbitrary geometry, allowing the inclusion of  coil features in the design process such as maximization of the magnetic field in a specific region or minimization of the resistance (power dissipation)  \cite{Poole_2007,NovelTMS}. Stream function IBEM is therefore a strategy well suited for use in electronics to design optimal transducers capable of covering the magnetic field range of AMR sensors with reduced power dissipation.

In this article, a stream function IBEM is applied to design a PCB coil that allows in-flight noise characterization of a low-frequency magnetic measurement system. This scheme has been used to produce a set of optimal transducers on two available surfaces of the multilayer PCB  where the magnetic diagnostic system is implemented. The coil has been designed to maximize the relevant magnetic field component of the corresponding AMR sensor while reducing power dissipation. The transducer proposed was also simulated, built and experimentally validated. Its remarkable performance allowed its implementation on the MELISA-III payload, which  will be integrated in a 6-Unit CubeSat platform by ISISpace.

This paper is structured as follows. Firstly, the stream function IBEM is outlined. Secondly, a brief overview is given of the main characteristics of the proposed magnetic measurement system. Next, the geometrical and functional requirements the coil transducer must meet are listed. Finally the coil prototype designed is introduced, with a description and discussion of the results obtained from the numerical simulations and experimental measurements.

\section{Methods}


\subsection{Stream function IBEM}
\label{section:IBEM}

The following is a brief outline of the method. For a more detailed description refer to the original works \cite{Poole_2007, Pissanetzky_1992}.

A stream function IBEM for coil design is a current density technique, in the sense that  it is based on the idea of wire arrangements wound so as to approximate continuous current distributions. In this type of approach, the desired current density is determined by solving an optimization problem formulated in terms of the stream function. In order to obtain the current density, the conducting surface of the coil, $S$, is discretized in a triangle mesh elements and $N$ nodes, which are located at each vertex of the element. The current density on the conducting surface can be represented as:
  \begin{equation}
\textbf{J}(\textbf{r})\approx\sum_{n=1}^N \psi_{n} \
\boldsymbol{\jmath}^{n}(\textbf{r}), \qquad \textbf{r} \in S, \label{bem131}
\end{equation}
where $[\psi_{1}, \psi_{2},\ldots, \psi_{N}]$ are the stream function nodal values used as optimization variables, and $\boldsymbol{\jmath}^{n}$ is the current element associated to the $n^{th}$-node \cite{Poole_2007}.

By using the current density model in Eq. \ref{bem131}, we can obtain the discretized expressions for the required physical magnitudes involved in the design problem. For instance, the magnetic field created by the coil at a given point can be written as:
\begin{equation}
\textbf{B}(\textbf{r})\approx\sum_{n=1}^N \psi_{n} \textbf{b}^n(\textbf{r}), \label{bembzbz}
\end{equation}
where $\textbf{b}^n$ is the magnetic induction vector produced by a unit stream function at the \textit{nth}-node \cite{Poole_2007}. Analogously, the amount of resistive power dissipated by the coil can be expressed as: 
\begin{equation}
P\approx \sum_{n=1}^N \sum_{m=1}^N \psi_{n} \psi_{m} R_{nm} , 
\end{equation}
where $R$ is the resistance matrix, an $N\times N$ sized fully symmetric positive definite matrix \cite{Poole_2007}.

In an IBEM these discrete models are used to pose the coil design as a convex optimization, with the solution being the optimal stream function. The final wire arrangement is achieved firstly by equally-spaced contouring of the stream function \cite{Brideson2002}, and secondly by connecting the resulting unconnected contour lines in series (which is by far the main source of error of this technique). The stream function IBEM is frequently used to generate coil windings with wires \cite{Rogowski2021}, but it can also be applied to variable or fixed-width copper tracks \cite{While2013}. In this article, the latter was preferred to produce a coil made from PCB tracks.

\subsection{Magnetic measurement system} \label{section:sensors}
\begin{figure}[h]
    \begin{center}
        \includegraphics[width=0.45\textwidth]{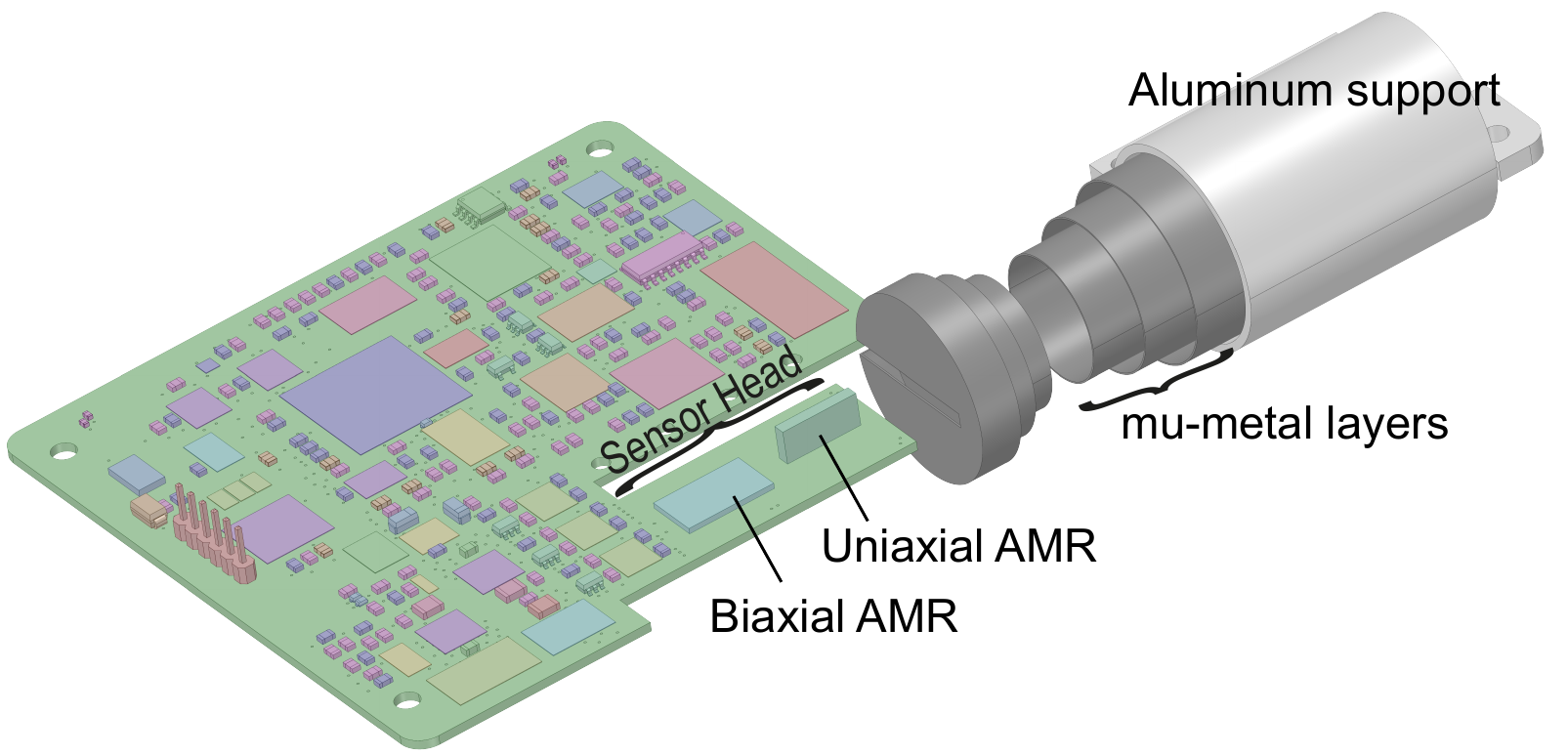} 
    \end{center}
    \caption{Melisa-III payload consisting of four-layer PCB  along with three mu-metal layers for the magnetic shielding and the aluminium support.} \label{fig:Payload}
\end{figure}

The magnetic measurement system incorporated on the MELISA-III payload consists of a four-layer PCB with a sensor head made up of single axis (HMC1001) and two-axis (HMC1002) AMR magnetometers \cite{HMCSensors}, along with the corresponding analog and digital circuits that acquire the sensor measurements, digitize them and communicate with the satellite's on-board computer.
Furthermore, the AMR sensors were placed within a cylindrical enclosure to shield the magnetometers from the environmental magnetic field \cite{Shielding},\cite{MateosCubeSat}, thus allowing in-flight noise characterization of the sensors at sub-millihertz frequencies. 
The shielding consisted of three concentric cylindrical mu-metal layers with a thickness of $0.254$ mm attached to the payload by a cylindrical aluminum structure with a thickness of $1.2$ mm. Each mu-metal layer was separated by a gap of $1.7$ mm, as shown in Fig. \ref{fig:Payload}.

\subsection{Transducer design}
\label{section:CoilDesign}
The transducer coil was designed to fit in the available area of the multilayer PCB after implementation of the analog and digital circuits.
More precisely,   the transducer was created taking into account the free surface of the two middle layers of the 
four-layer PCB structure in the sensor head region, which allows a reduced distance between the transducer and the sensor, thereby improving the strength and uniformity of the magnetic field over the magnetometers.

\begin{figure}[h]
    \begin{center}
        \subfigure[] {\label{fig:Esq_3D}\includegraphics[width=0.475\textwidth]{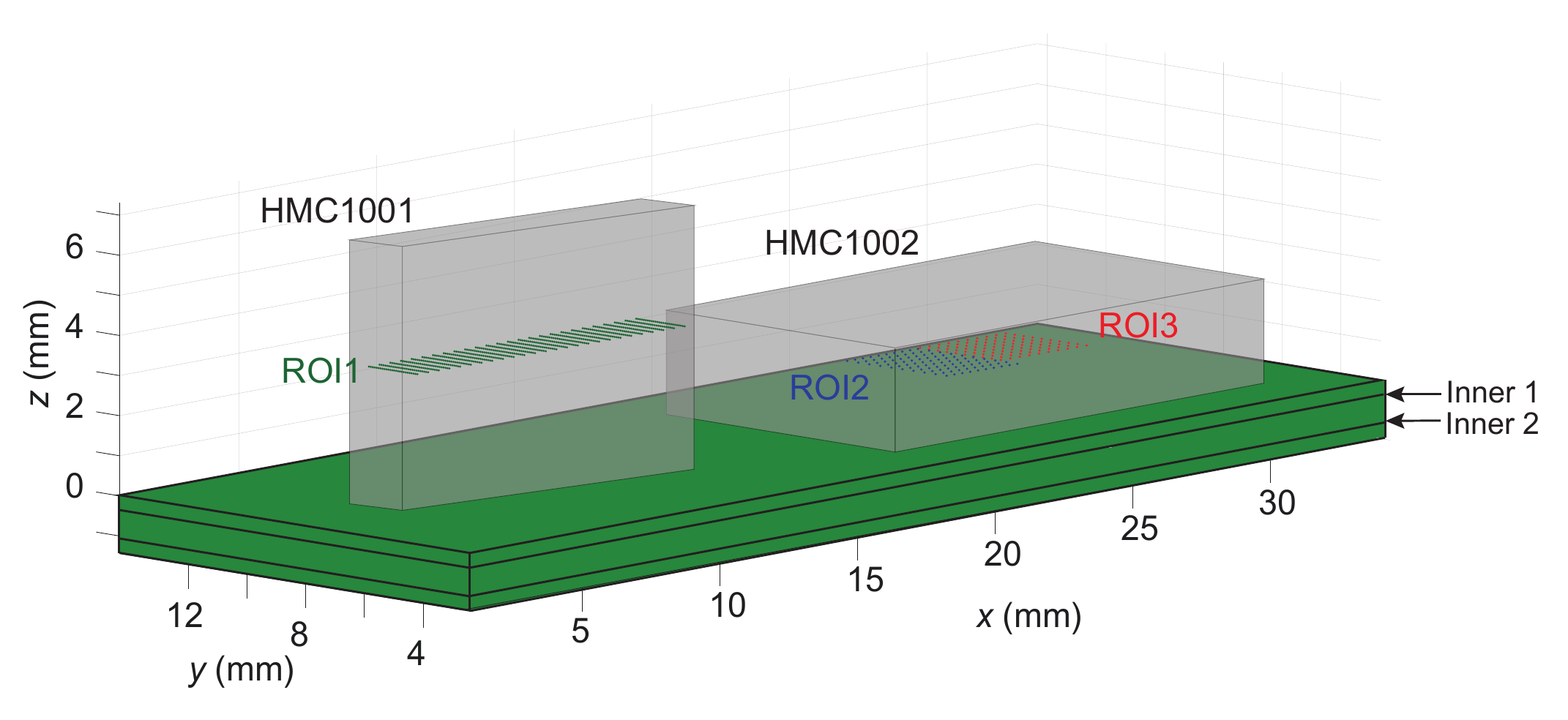}} %
        \subfigure[] {\label{fig:Esq_3D_XZ}\includegraphics[width=0.475\textwidth]{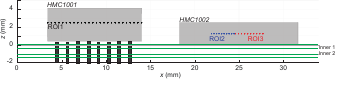}} %
    \end{center}
    \caption{a) 3D schematic diagram and b) $xz$ projection showing the sensor head region of MELISA-III payload with the AMR magnetometers. The transducer coil is designed in the two available areas (Inner 1 and Inner 2) of the two middle layers with the objective of maximizing $B_Z$ at ROI1, $B_x$ at ROI2 and $B_y$ at ROI3.} 
    \label{fig:PCB_Esquema}
\end{figure}

The transducer was designed in two selected areas (each in one of the two inner layers of the four layers PCB), with having an initial feasible surface of $33.35$ mm $\times$ $15.7$ mm. Hereinafter, these two available PCB areas will be referred to as Inner 1 and Inner 2, they are located $0.36$ mm and $1.07$ mm, respectively, below the top surface of the multilayer PCB ($z=0$ mm) where the AMR magnetometers are placed (see Figs. \ref{fig:Esq_3D} and 2(b)).

 The sensor head is made up of a uniaxial (HMC1001) and a biaxial AMR (HMC1002) \cite{HMCSensors}, where the $z$-axis is the sensitive direction for HMC1001, and $x$ and $y$ are the sensitive axes for  the biaxial HMC1002,  as shown in Figs. \ref{fig:Esq_3D} and 2(b). 
 
 According to this spatial distribution, the desired transducer coil must create a magnetic field for which the $z$ component is maximum in the HMC1001 region,  whereas the $B_x$ and $B_y$ produced must also be maximum in the region occupied by HMC1002. To this end, three regions of interest (ROIs) were set for  maximization of the different magnetic field components. 
The first region of interest ROI1 was located within the HMC1001 sensor at $z=2.5$ mm over the top layer surface ($z=0$ mm) of the multilayer PCB with the aim of achieving a maximum $B_z$. ROI2 and ROI3 were placed inside the HMC1002 sensor at $z=1.25$ mm over the PCB surface with the objective of maximize the $B_x$ and $B_y$ components, respectively. 
 It is worth stressing that the exact sensing regions of each magnetometer is unknown, nonetheless this placement of the three ROIs is a plausible choice according to the sensor characteristics  \cite{HMCSensors}.

\begin{figure}[h]
    \begin{center}
        \subfigure[] {\label{fig:Dim_Inner1}\includegraphics[width=0.45\textwidth]{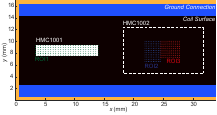}} %
        \subfigure[] {\label{fig:Dim_Inner2}\includegraphics[width=0.45\textwidth]{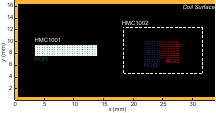}} %
    \end{center}
    \caption{Available surfaces for the coil design at: a) Inner 1 and b) Inner 2. The ground tracks  in Inner 1 are depicted as two blue bands of $2$ mm width. The HMC1001 AMR sensor is showed as a white rectangle to illustrate the region where coil tracks are excluded due to the drill holes. The HMC1002 sensor location is showed as a dotted rectangle. (The ROIs are shown for sake of clarity, but it is worth stressing that Inner 1 plane is at $z=-0.36$ mm, Inner 2 plane is at $z=-1.07$ mm whereas  ROI1 is the HMC1001 sensor at $z=2.5$ and ROI2 and ROI3 are inside the HMC1002 sensor at $z=1.25$.)} 
    \label{fig:PCB_Surface}
\end{figure}

Furthermore, although the initial sensor head  surface is $33.35$ mm $\times$ $15.7$ mm per layer, the actual surface available for the transducer design must also meet the following geometrical  restrictions: 

\begin{enumerate}[i)]
    \item Two ground tracks are implemented in the Inner 1 area  (blue region in Fig, \ref{fig:Dim_Inner1}). Consequently,  these $2$ mm wide zones have not been considered as part of the available surface for the design.
   
    \item HMC1001 is a through-hole component and so the area of insertion holes cannot be used in the design. Two rectangular regions of $10.6$ mm $\times$ $1.8$ mm in both layers, Inner 1 and Inner 2 have not been considered as part of the available surface in the design, these areas from which the coil tracks are excluded are shown in Figs. \ref{fig:Dim_Inner1} and \ref{fig:Dim_Inner2}.
 
 \end{enumerate}

In addition, the  conditions for fabrication  of the PCB introduce the following requirements:
\begin{itemize}
    \item Minimum width for the coil tracks of $0.15$ mm.
    \item Minimum distance between coil tracks of $0.15$ mm.
    \item PCB copper thickness of $0.035$ mm.
\end{itemize}
In order to achieve the highest number of coil tracks (and so the highest magnetic field generated by the transducer), the track width is considered as the minimum value prescribed by the fabrication conditions 

The transducer was also designed based on the following functional requirements:
\begin{itemize}
    \item  It must generate an average bias magnetic field with the desired direction in the corresponding ROI  of each sensor of the order of $4\mu$T for $B_x$ and $B_y$ components and $10 \mu$T for $B_z$. Since the magnetic field range for MELISA-III is $\pm 10\:{\rm \mu T}$ and the electronic signal conditioning circuit is common for the three axes, the proposed magnetic field generated by the PCB coil will be sufficient to effectively characterize the whole magnetic measurement system.
 
    \item Low minimum coil resistance is required to reduce the power dissipation, so $R$ has been set to be lower than $10 \,\Omega$.
    
\end{itemize}

The different performance, geometrical and functional requirements listed above allow consideration of the coil design as a multi-objective constrained optimization problem, the solution of which can be used to produce the desired PCB tracks arrangement.

\subsection{Numerical validation}

COMSOL Multiphysics software (COMSOL AB, Stockholm, Sweden) \cite{COMSOL} was used to numerically validate the designed multilayer PCB transducer,  which  have proved to be an effective tool to simulate the performance of coils in electronics \cite{Yang2021,Ding2022}. A 3D model of the coil was created at the COMSOL geometry section and the \textit{AC/DC Module} with the \textit{Magnetic Fields} physics and \textit{Stationary} study for an input coil current of $10$ mA were employed to perform an analysis of the magnetic field generated at each of the three ROIs. The resistance and the inductance of the transducer coil were also simulated in the same manner.

\subsection{Experimental validation}
Magnetic measurements were carried out with the MELISA-III payload to verify the theoretical and simulated results of the PCB coil designed. For these measurements, the payload was calibrated with a triaxial Helmholtz coil, with the current-to-field ratio  previously estimated using an absolute optically pumped atomic magnetometer.

\begin{figure}[h]
    \begin{center}
        \includegraphics[width=0.45\textwidth]{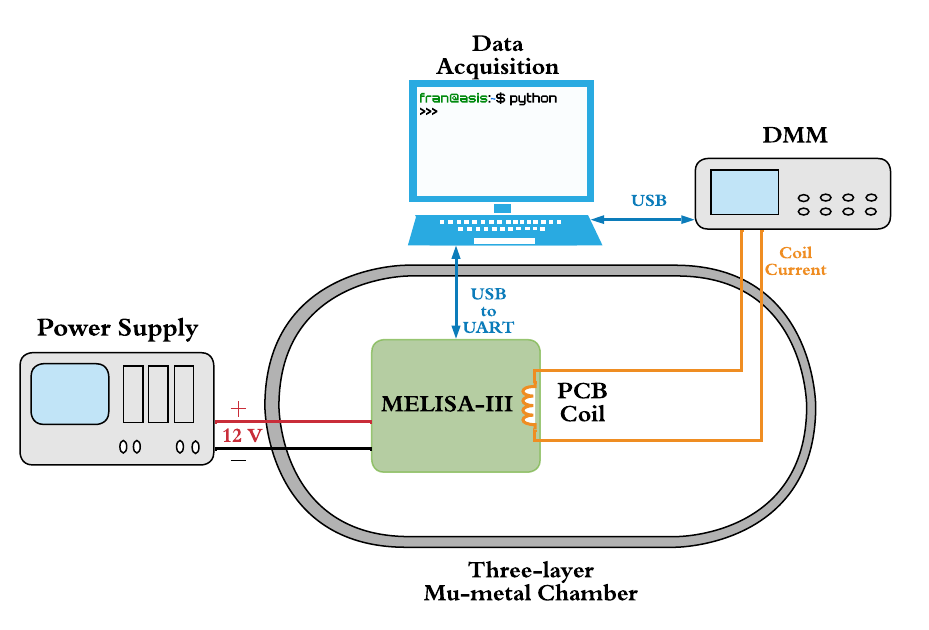} 
    \end{center}
    \caption{Block diagram of the experimental set-up employed for validating the PCB transducer coil.} \label{fig:PCBCoilSetup}
\end{figure}

The payload was allocated in a magnetically shielded environment to screen out the Earth's magnetic field and magnetic sources in the lab. The measurements were performed both with and without MELISA-III's three-layer cylindrical mu-metal shielding (see Fig.\ref{fig:Payload}). This allowed assessment as to whether the magnetic field induced by the PCB transducer on the sensor location is modified due to the proximity of the high-permeability mu-metal alloy. The current applied to the coil is acquired by an external high-accuracy 7.5 digit multimeter (DMM). The corresponding block diagram of the experimental set-up employed to validate the PCB transducer coil is shown in Fig. \ref{fig:PCBCoilSetup}.

\section{Results}
\subsection{PCB Transducer}
\begin{figure}[h]
    \begin{center}
        \subfigure[] {\label{fig:Inner1s}\includegraphics[width=0.45\textwidth]{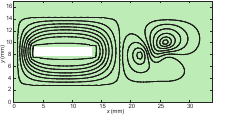}} %
        \subfigure[] {\label{fig:Inner2s}\includegraphics[width=0.45\textwidth]{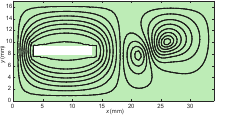}} %
    \end{center}
    \caption{ Stream function contour lines of the a) Inner 1 layer and b) Inner 2 layer.} 
    \label{fig:StreamF_Contour}
\end{figure}

By using the IBEM approach described in section \ref{section:IBEM} along with the requirements presented in section \ref{section:CoilDesign}, the design problem is posed as a constrained optimization from which the desired stream function (and so current density) can be obtained.  Figs. \ref{fig:Inner1s} and \ref{fig:Inner2s} show the stream lines produced by contouring the stream function solution at the Inner 1 and Inner 2 layers, respectively. 

\begin{figure}[h]
    \begin{center}
        \subfigure[] {\label{fig:Inner1}\includegraphics[width=0.45\textwidth]{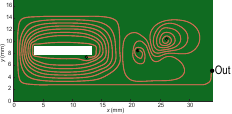}} %
        \subfigure[] {\label{fig:Inner2}\includegraphics[width=0.45\textwidth]{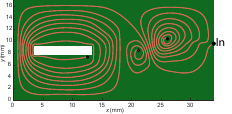}} %
    \end{center}
    \caption{ PCB track distribution of the coil transducer at a) Inner 1 layer and b) Inner 2 layer. Input (In) and output current (Out) are set on the Inner 2 and Inner 1, respectively.  
       The connections between layers are depicted by black circles (crosses within the circle represent an input current into the connection path whereas dots represent an output current).}
    \label{fig:PCB_Coil}
\end{figure}

By connecting these contour lines in series, we obtain the final track pattern of the coil depicted in Fig. \ref{fig:PCB_Coil}. It is a six lobed coil, where Fig. \ref{fig:Inner1} and Fig. \ref{fig:Inner2} correspond to the PCB track arrangement of the Inner 1 and the Inner 2 layers, respectively. A higher density of PCB wires can be appreciated in the PCB regions closest to the sensors, $i.e.$ where the highest magnetic fields are expected. In addition, it can be seen how the ground connection surface in Inner 1 (Fig. \ref{fig:Dim_Inner1}) as well as the drill hole regions of the HMC1001 in both layers (Fig. \ref{fig:PCB_Surface}) are free of PCB tracks, as expected since they were defined as excluded regions in the design process (section \ref{section:CoilDesign}).

The input and output current connections are set on the Inner 2 and Inner 1 layers respectively, where the current flow is guaranteed by three connection paths between layers (black circles in Fig. \ref{fig:PCB_Coil}).

\subsection{Numerical validation} \label{section:comsol1}

\begin{figure}[h]
    \begin{center}
        \subfigure[] {\label{fig:BX}\includegraphics[width=0.45\textwidth]{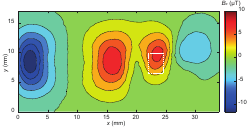}} %
        \subfigure[] {\label{fig:BY}\includegraphics[width=0.45\textwidth]{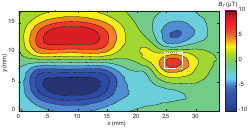}} %
        \subfigure[] {\label{fig:BZ}\includegraphics[width=0.45\textwidth]{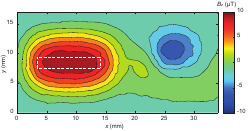}} %
    \end{center}
    \caption{ Colour-code map of  a) $B_x$ component of the magnetic field at $z = 1.25$ mm, b) $B_y$ component of the magnetic field at $z = 1.25$ mm and c) $B_z$ component of the magnetic field at $z = 2.5$ mm. The ROIs are represented by white rectangular dotted lines.}
    \label{fig:Bfied_TEO}
\end{figure}

Figs. \ref{fig:BX}, \ref{fig:BY} and \ref{fig:BZ} show the colour-coded maps of the three components $B_x$, $B_y$ and $B_z$ of the magnetic field produced by the transducer coil in Fig. \ref{fig:PCB_Coil} when simulated with COMSOL multi-physics using an input coil current of 10 mA. The fields $B_x$ and $B_y$ are shown on the $xy$-plane at $z = 1.25 \ \rm{mm}$, whereas $B_z$ is depicted on the $xy$-plane at $z = 2.5$ mm.
It is worth noting that the maximum values of $B_x$ and $B_y$ occur at the biaxial sensor locations, Figs. \ref{fig:BX} and \ref{fig:BY} respectively,  whereas the maximum strength of $B_z$ occurs in the uniaxial sensor region, Fig. \ref{fig:BZ}. Consequently in all cases the maximum value of the desired magnetic field component coincides with the  sensitive axis of the corresponding magnetometer. In addition, the B-field strengths meet the initial requirements to efficiently unbalance the Wheatstone bridge up to the maximum magnetic field range of the MELISA-III payload, i.e. $\pm 10\:{\rm \mu T}$.

\begin{figure}[h]
    \begin{center}
        \subfigure[] {\label{fig:BXX}\includegraphics[width=0.45\textwidth]{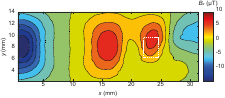}} %
        \subfigure[] {\label{fig:BYY}\includegraphics[width=0.45\textwidth]{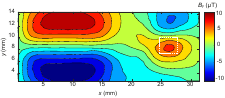}} %
        \subfigure[] {\label{fig:BZZ}\includegraphics[width=0.45\textwidth]{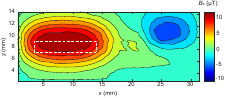}} %
    \end{center}
    \caption{ Colour-code map of a) $B_x$ component of the magnetic field at $z = 1.25$ mm, b) $B_y$ component of the magnetic field at $z = 1.25$ mm and c) $B_z$ component of the magnetic field at $z = 2.5$ mm considering the $\mu$-metal cylindrical shielding. The ROIs are represented  by white rectangular dotted lines.}
    \label{fig:mumetal_Bfied_TEO}
\end{figure}

These results indicate that the designed transducer fulfills the initial requirements (sections \ref{section:CoilDesign}). Nonetheless, for an accurate evaluation of the magnetic measurement system the shielding effect of the three cylindrical mu-metal layers must also be considered in the analysis. To this end,  the bias magnetic field produced by the coil within the cylindrical shielding has also been simulated using COMSOL multi-physics. Figs. \ref{fig:BXX}, \ref{fig:BYY},\ref{fig:BZZ} illustrate the colour-coded map of the $B_x$, $B_y$ and $B_z$ in the equivalent $xy$-planes considering the influence of the magnetic shielding. It can be seen that the maximum values of $B_x$, $B_y$ and $B_z$ occur again at the sensors locations. The presence of the magnetic shielding does not significantly affect the magnetic field strength at the ROIs, as it can be seen in Figs.  \ref{fig:Bfied_TEO}, \ref{fig:mumetal_Bfied_TEO} and in Table \ref{tab:TransProp}. The B-field strengths are also met in this case in the range of the $4-10\:{\rm \mu T}$.

In addition, the maximum magnitudes of $B_z$, $B_x$ and $B_y$ at  ROI1, ROI2 and ROI3 respectively are detailed in Table \ref{tab:TransProp}, where it is noticeable that the COMSOL results are lower than those  obtained in the theoretical analysis. This may be partially explained due to the series connection that must be performed in the coil after the stream function contouring process, as described in section \ref{section:IBEM}. Nonetheless, it can be seen how the maximum values are withing the desired B-field range (between $4$ and $10\:\mu$T) mentioned in section \ref{section:CoilDesign}.


\begin{table*}[h!]
    \centering
    \caption{Transducer properties along with maximum and average magnetic field components produced by the transducer at each ROI ($Z$, $X$ and $Y$ at ROI1,ROI2 and ROI3, respectively) for $I$ = 10 $\text{mA}$. }
    \label{tab:TransProp}
    \begin{tabular}{cccccc}
        \hline
         & Theoretical     & COMSOL    & COMSOL  & Experimental & Experimental   \\
         &      &      & $\mu$-metal &  &  $\mu$-metal \\
        \hline
        Resistance ($\Omega$)            &    3.0        &   3.0       &   3.0              &    5.5   &   5.5    \\
        \hline
        $B_{x{\rm max}}$ ($\mu$T)              &    8.7        &   8.3       &   8.1              &      -    &   -        \\
        $B_{y{\rm max}}$ ($\mu$T)              &    7.2        &   7.6       &   7.3              &      -    &   -        \\
        $B_{z{\rm max}}$ ($\mu$T)              &    10.8       &   10.9      &   12.7             &      -    &   -        \\
        \hline
        $\bar{B}_x$ ($\mu$T)             &    6.7        &   6.0       &   5.7              &    4.6   &    4.3     \\
        $\bar{B}_y$ ($\mu$T)             &    5.3        &   5.6       &   5.5              &    4.2   &    4.0      \\
        $\bar{B}_z$ ($\mu$T)             &    10.0       &   9.9       &   11.8             &    10.5  &    12.3     \\
        \hline
    \end{tabular}
    
\end{table*}

\subsection{Experimental validation}

\begin{figure}[h]
    \begin{center}
        \label{fig:foto}\includegraphics[width=0.45\textwidth]{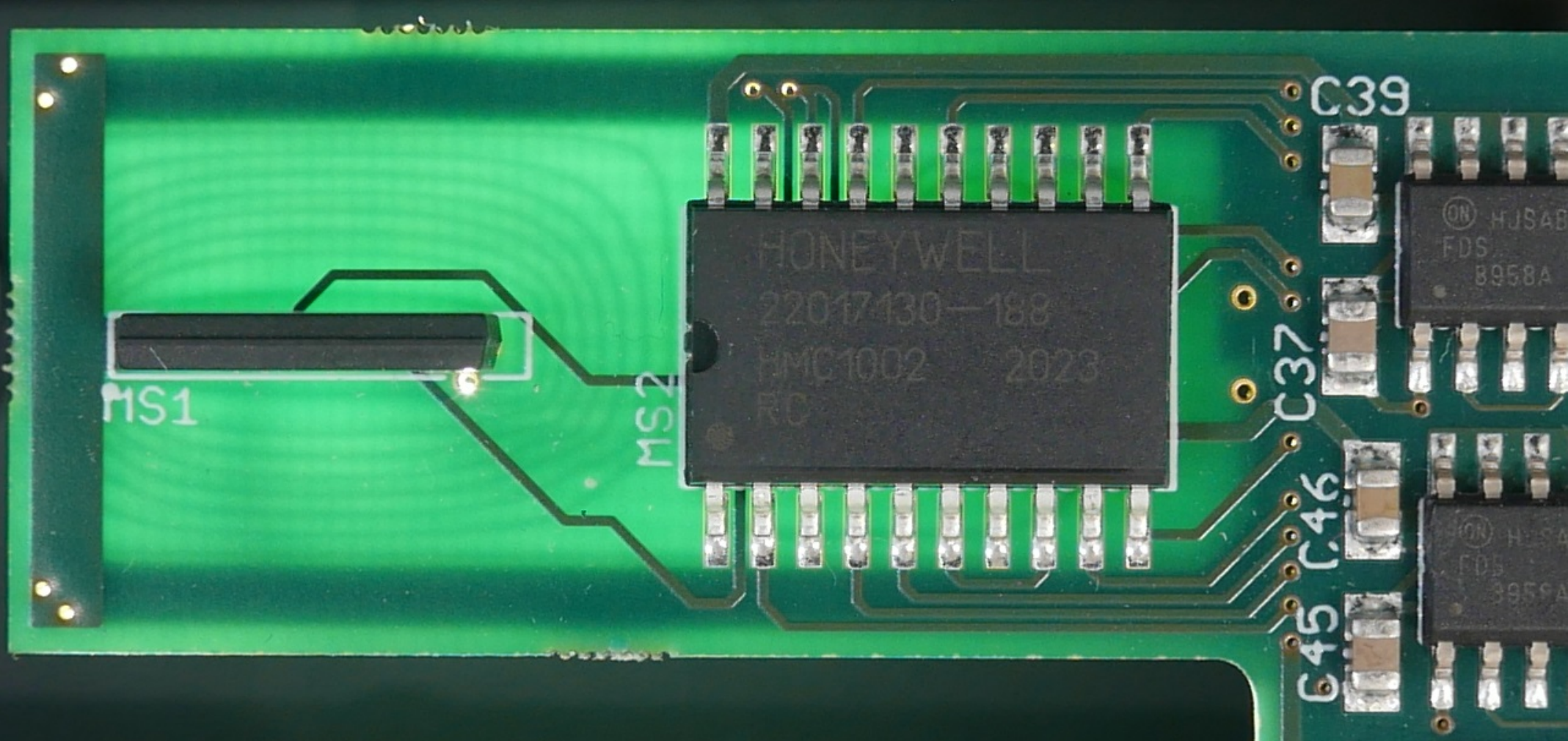} 
    \end{center}
    \caption{Photographs of the Melisa-III payload. The figure depicts the sensor head region with the AMR sensors and the coil designed.} \label{fig:Photo}
\end{figure}

The sensor head of the magnetic measurement system fabricated can be seen in Fig. \ref{fig:Photo}, where the PCB tracks of the coil may also be partially distinguished.  The resistance of this transducer and the relevant maximum and average magnetic fields produced in each ROI are presented in Table \ref{tab:TransProp}, where these experimentally measured values can be compared to those theoretically derived from the IBEM or simulated by using COMSOL.

The measured resistance of the prototype transducer coil in Fig. \ref{fig:Photo} was found to be in accordance with the simulated and theoretical values, which confirms the accuracy of modelling coils in this way. It should be noted that the resistance may be influenced by the copper thickness tolerance, meaning that the transducer track thickness in some regions may be less than the nominal value indicated by the manufacturer, leading to  higher resistance measurements. The thickness may be as low as $0.024$ mm (instead of the $0.035$ mm considered in the simulation), which may explain the  discrepancy between theoretical and simulated resistance with respect to the experimentally obtained value.

\begin{figure}[h!]
    \begin{center}
        \subfigure[] {\label{fig:Bx-I}\includegraphics[width=0.45\textwidth]{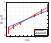}} %
        \subfigure[] {\label{fig:By-I}\includegraphics[width=0.45\textwidth]{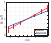}} %
        \subfigure[] {\label{fig:Bz-I}\includegraphics[width=0.45\textwidth]{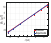}} %
    \end{center}
    \caption{ Magnetic field-current plot for the MELISA-III payload with the $\mu$-metal shielding.  a) $B_x$ , b) $B_y$,   and c) $B_z$ components of the magnetic field. The red curves depict the average value of the relevant component obtained from Comsol simulation over   a) ROI2, b) ROI3 and c) ROI1. The blue curves show the measured values. The errors in the red curve are due to the standard deviation of the field component over the corresponding ROI. The errors in the experimental curves are of the order of nano tesla and are associated with the uncertainty of the measuring instrument and the resolution of the MELISA-III. }
    \label{fig:B-I_res}
\end{figure}

On the other hand, the measured (blue line) and numerically calculated (red line) values of each magnetic field component considering the mu-metal shielding are shown for different current intensities through the coil transducer in Fig. \ref{fig:B-I_res}. The simulated fields are found as the average value of the relevant component in the corresponding ROI. In order to evaluate the uncertainty, the standard deviation of the field component over the corresponding ROI has also  been included for each current.  The errors in the experimental curves are of the order of nano tesla and so negligible.

It is worth noting that there is an appreciable difference between the average values of the simulated $B_x$ and $B_y$ components and those measured by the HMC1002 sensor, as can be seen in Figs. \ref{fig:Bx-I}, \ref{fig:By-I} respectively. 
This fact indicates that ROI2 and ROI3 do not completely match the most sensitive regions of the biaxial magnetometer, nonetheless the three  regions of interest chosen proved to be a representative approximation to design coils able to successfully characterize the entire magnetic measurement system. In addition,  the deviation between simulations and measurements may be justified, in part by the fact that the final chip height may change from the one originally considered in Fig. \ref{fig:PCB_Esquema} because of the deviations introduced by soldering (up to several tenths of a millimetre in the $z$ direction).The influence of these sources of error is not significantly high in the case of the $z$ component of the magnetic field,  Fig. \ref{fig:Bz-I}, as the average value of the simulated $B_z$ over  ROI1 offers a good approximation of the HMC1001 sensor measurement.   Regardless, all the B-field components produced by the transducer coil in both magnetometers fully satisfy the initial functional and performance requirements.

\section{Conclusion}

This article describes   a multilayer PCB transducer to ensure stable magnetic conditions in a technology demonstrator of a magnetic diagnostic system for future space-borne GW detectors.
The design method was based on a stream function IBEM adapted to produce multilayer PCB coils, allowing consideration of complex coil geometries and 
prototyping of different functional constraints such as minimum power dissipation in the design process.
The performance of the transducer presented  was simulated and experimentally evaluated. The theoretical, numerical and measurement values were found to be in good agreement, fulfilling in any case the performance requirements and efficiently controlling the magnetic field in the magnetic measurement system, even when the sensors are placed within a magnetically shielded enclosure made of three cylindrical mu-metal layers.
Furthermore, the geometry-independent nature of this coil design approach, its versatility for consideration numerous performance requisites and  its direct implementation using conventional PCB technology make the stream function IBEM   a promising technique to produce specific PCB coils in electronics.

\section*{Acknowledgment}
This work has been co-financed by the 2014-2020 ERDF Operational Programme and by the Department of Economy, Knowledge, Business and University of the Regional Government of Andalusia (Project reference: FEDER-UCA18-105867 and FEDER-UCA18-107380) and PAIDI-2020 Programme (Project reference: P18-FR-2721).  The MELISA-III experiment will be embarked on the first mission of the H2020 IOD/IOV program of the EU.

\section*{Rights Statement}
\textcopyright{} 2022 IEEE. Personal use permitted. The published version is available at: \url{https://doi.org/10.1109/tim.2022.3217843}. This manuscript is the accepted version of the article published in \textit{IEEE Transactions on Instrumentation and Measurement}.

\printbibliography 


\end{document}